\documentstyle[12pt]{article}
\def\eslt{E\llap/_T}
\def\to{\rightarrow}
\def\Re{{\cal R \mskip-4mu \lower.1ex \hbox{\it e}}\,}
\def\Im{{\cal I \mskip-5mu \lower.1ex \hbox{\it m}}\,}

\def\te{\tilde e}
\def\tl{\tilde l}

\def\tg{\tilde g}

\def\tnu{\tilde\nu}
\def\tell{\tilde\ell}
\def\tq{\tilde q}
\def\tb{\tilde b}
\def\tt{\tilde t}
\def\tw{\widetilde W}
\def\tz{\widetilde Z}
\def\tf{\tilde f}

\def\alt{\stackrel{<}{\sim}}
\def\agt{\stackrel{>}{\sim}}

\begin {document}
\begin{flushright}
UH-511-803-94\\
September, 1994
\end{flushright}
\begin{center}
\vglue 3.5cm
{\large \bf LOOKING BEYOND THE STANDARD MODEL\footnote{Invited Talk,
presented at the Physics in Collision Conference, Tallhassee, Florida, June,
1994}}\\
\vspace{6.0ex}
{\large XERXES TATA}\\
\vspace{1.5ex}
{\large \it Department of Physics and Astronomy, University of Hawaii,\\
Honolulu, HI 96822, U.S.A}\\
\vspace{1.5ex}
\vspace{1.0cm}
\begin{abstract}
Within the framework of the Standard Model,
the scale of electroweak symmetry breaking is
unstable to radiative corrections. We discuss two broad classes
of models of new physics (one with a strongly
interacting and the other with a perturbatively
coupled electroweak symmetry breaking sector) in which
this stability is restored.
After reviewing experimental constraints on these,
we discuss the implications of these types of models for experiments, both
at currently operating colliders as well as the next generation of colliders
under consideration for construction. Other extensions of the Standard Model
are briefly alluded to.
\end{abstract}
\end{center}

\newpage

\section{The Status Of The Standard Model}
\subsection{Experimental Report Card.}

It has become standard practice to begin such a review by showering praise
upon the Standard Model (SM), which has indeed been spectacularly successful
in accommodating a wide variety of experimental data. As discussed by
Olchevski\cite{OLCH} at this meeting, the results from experiments at LEP
shown in Table 1 appear to be in remarkable agreement with the SM.

\begin{table}[htb]
\centering
\caption[]{\small Summary of experimental electroweak measurements and
corresponding SM fits from
Ref. \cite{OLCH}. The last entry is from Ref. \cite{FERO}
The entry in the last column represents the difference
between the measured and fitted values, expressed as the number of standard
deviations.}
\bigskip
\begin{tabular}{|c|c|c|c|} \hline
Observable & Measurement & SM fit & Pull \\ \hline
{\bf LEP} &  &  &\\
$M_Z$ (GeV) & $91.1895\pm 0.0044$ & 91.192 & 0.6 \\
$\Gamma_Z$ (GeV) & $2.4969 \pm 0.0038$ & 2.4967 & 0.1 \\
$\sigma^0_h$ ($nb$) & $41.51 \pm 0.12$ & 41.44 & 0.6 \\
$R_{\ell}$ &$20.789 \pm 0.04$ &20.781 & 0.2 \\
$A^{0,\ell}_{FB}$ & $0.0170 \pm 0.0016$ & 0.0152 & 1.1 \\ \hline
$A_{\tau}$($\tau$ pol.) & $0.150 \pm 0.010$ & 0.142 & 0.8 \\
$A_e$($\tau$ pol.) & $0.120 \pm 0.012$ & 0.142 & 1.8 \\ \hline
$R_b$ & $0.2208 \pm 0.0024$ & 0.2158 & 2.0 \\
$R_c$ & $0.170 \pm 0.014$ & 0.172 & 0.1 \\
$A^{0,b}_{FB}$ & $0.0960 \pm 0.0043$ & 0.0997 & 0.8 \\
$A^{0,c}_{FB}$ & $0.070 \pm 0.011$ & 0.071 & 0.1 \\
$\sin^2\theta^{lept}_{eff}$ from $q\bar{q}$ charge asymm. & $0.2320\pm
0.0016$ & 0.2321 & 0.1 \\ \hline
{\bf $p\bar p$ and $\nu N$} &   &  & \\
$M_W$ (GeV) & $80.23 \pm 0.18$ & 80.31 & 0.4 \\
$1-\frac{M_W^2}{M_Z^2}$ ($\nu N$) & $0.2256 \pm 0.0047$ & 0.2246 & 0.2 \\
\hline
{\bf SLC} &  &   & \\
$\sin^2\theta^{lept}_{eff}$ pol. asymm. & $0.2292 \pm 0.0010$ & 0.2321 & 2.7
\\
\hline
\end{tabular}
\end{table}

Of the
twelve quantities measured at LEP, only $R_b = \Gamma(b\bar b)/\Gamma(had)$
deviates from the SM fit by $2\sigma$. Since there is a
probability of about 5\% that any measurement will yield a $2\sigma$
deviation, the probability that the twelve independent LEP measurements will
all agree to within $2\sigma$ is about 54\% (or slightly larger, depending
on how one counts input parameters). The conservative would thus say
that there is essentially a
50-50 chance that at least one of the LEP measurements
would show such a deviation, and so, would conclude that there is no
problem whatsoever at the present time,
pointing also to the $M_W$ and neutrino scattering
measurements in Table 1 which are in agreement (within larger errors)
with the SM.

The radical scientist\footnote{The extremist
would counter the conservative as follows: Any theory that agrees with
all the experimental data has to be erroneous, since at any given time some
of the experimental data are wrong. We will, of course, disregard such
an extreme view.}, on the other hand, would point
to the polarization
asymmetry measurement\cite{FERO} at the SLC and conclude that this,
together with the measurement of $R_b$ and $A_e$ (from $\tau$ polarization)
at LEP, provide evidence for physics beyond the SM. To bolster these
arguments, one could refer to the model-independent analysis of
\cite{PESKIN} where the (oblique) corrections from
any new physics, assuming that the new physics scale is well-separated
from $M_Z$, are parametrized in terms of just three parameters S, T and U
which are all zero in the SM. Indeed the radical would
emphasize \footnote{If the new physics scale is
close to $M_Z$, the corrections can no longer be parametrized in terms
of just S,T and U. Fits including additional parameters would
then make the S-parameter consistent with zero, within the resulting larger
error\cite{BURGESS}.}
that $S<0$ ($2\sigma$)\cite{LANG}.

Are there other hints of deviation from the SM? We have heard from
Kuhlmann\cite{KUHLMANN} that, generally speaking, perturbative QCD appears
to be working well. The only aberration reported
appears to be an excess of $p_T < 30$
GeV photons in direct photon production at CDF (which suggests
a $k_T$ smearing of the initial state). The CDF
measurement\cite{SHOCHET} of the cross section for direct $\psi'$ production,
which is 20-30 times the present theoretical estimates, may also be worth
watching. Whether this measurement (if it holds up) signals an unthought-of
production mechanism, or something deeper, only time will tell. On the
neutrino front, the recent observations of the appearance
of $\bar{\nu}_e$ in the LSND experiment\cite{LSND},
or the zenith angle dependence
of the $\nu_{\mu}$ to $\nu_e$ ratio in the Kamiokande experiment\cite{KAM}
both point to neutrino mixing, and hence, a mass for neutrinos. While
this can be readily accommodated within the SM framework, the structure
of the neutrino mass matrix may help further our understanding of fermion
masses, or perhaps, provide clues about physics (new intermediate scales,
see-saw neutrino masses, or something else) at higher scales.
To my
knowledge, the only other ``clear discrepancy'' is an old one; {\it viz.}
the lifetime of orthopositronium\cite{ORTHO} is too short; if confirmed,
this would
be really revolutionary as it would signal a departure from QED!

\subsection {Theoretical Report Card.}
While it seems fair to say that the SM is in remarkable
agreement with experiment, it leaves
many items on the theorist's wish-list unexplained.

\begin{itemize}

\item The replication of
generations and the (perhaps, related issue of) patterns of fermion masses
(strongly
constrained by the absence of flavour-changing neutral currents(FCNC))
remains unexplained.

\item The choice of the gauge group and the values of the
three gauge couplings remain arbitrary.

\item There is no explanation (although there is a beautiful
parametrization) for the origin of CP violation.

\item The dynamics of electroweak symmetry breaking
(EWSB) is completely unknown.

\item Finally, and most strikingly, gravitation is
not incorporated.

\end{itemize}

There are no good answers to most of these issues. It is quite likely that
an understanding of these will come from knowledge of physics at very
high (and as yet unexplored) energy scales. The Grand Unification hypothesis
is a beautiful idea which, by enlarging the gauge group and particle
multiplets, leads to testable predictions for the relative strengths of the
three gauge couplings, and also for certain ratios of fermion masses. The
precise measurements of the gauge couplings at LEP, it is by now well
known\cite{UNIF}, are not compatible with the simplest Grand Unified model
based on SU(5). This can be ``fixed up'', for instance, by introducing new
scalars with appropriate lepto-quark quantum numbers; this modifies the
gauge coupling evolution so as to restore unification but is obviously
completely {\it ad hoc}. It is, however, very interesting that the measured
values of
gauge couplings are compatible with Grand Unification within the context of
the minimal model of low energy supersymmetry  which, as we will discuss later,
is introduced for very different reasons.

Despite the fact that the new physics of Grand Unified Theories (GUTs) is
all at an energy scale $ > 10^{15}$ GeV, GUT models make striking
predictions, the most generic of which is the instability of the proton.
Also, in all but the simplest models $B-L$ is not conserved, and $n\bar{n}$
oscillations are possible. Finally, the multiplet structure of models with a
gauge group larger than SU(5) includes singlet neutrinos, so that
non-vanishing neutrino masses and non-trivial neutrino mixing patterns may
be anticipated. We will not review the stringently constrained
phenomenology of GUT baryon number
violation here but refer the reader to the literature\cite{BARYON}.
Implications of neutrino masses and mixing for the solar and atmospheric
neutrino anomalies\cite{SUZ} as well as for neutrinoless $\beta\beta$
decay\cite{BETA}
have been discussed at this
meeting and will not be repeated. We will also not
discuss the implications (or non-implications) of GUTs for the observed
matter anti-matter asymmetry in the universe\cite{KOLB}, but turn instead
to EWSB in the SM.

Within the framework of the SM, EWSB is realized by introducing a doublet
of spin zero fields which acquire a vacuum expectation value, resulting in
the spontaneous breakdown of $SU(2) \times U(1)$.
The signature of this mechanism is an elementary
spin zero particle,
the Higgs boson, in the physical spectrum. Perturbative unitarity arguments
dating back twenty years\cite{PERT} suggest that the Higgs boson cannot be much
heavier than 700-800 GeV, provided, of course, that perturbation theory
is valid. An even stronger bound, $m_H \alt 220$ GeV can be
obtained\cite{SELF}
by requiring that the running Higgs self-coupling remains perturbative
all the way up to the GUT scale\footnote{This is also the origin of
the upper bound
of 140-150 GeV on the light Higgs boson in supersymmetric
models\cite{SUSYHIG}.}.

Unlike spin-$\frac{1}{2}$ and spin-1 particles whose masses can be
protected by chiral and gauge symmetries, respectively, there is no known
symmetry that protects the mass of a spin-0 particle in a generic quantum
field theory. Formally, this manifests itself as a quadratic divergence in
one-loop quantum corrections to the Higgs boson mass, when these are
computed using the SM framework.
These loop integrals
ought to be cut off at a scale $\Lambda$ beyond which the SM is
invalid, either because new degrees of freedom ({\it e.g.} additional
fields in GUTs) not included
in the SM manifest themselves, or due to form factor effects.
We may thus expect,
$\delta m_H^2 \sim g^2\Lambda^2$,
where the dimensionless coupling $g \sim O(1)$. We see that if $\Lambda \sim
M_{GUT}$ (or $M_{Planck}$), a Higgs mass below $\sim 800$~GeV
can only be obtained
by adjusting the bare Higgs mass term, order by order in perturbation theory,
with uncanny precision. Although there is nothing logically wrong with
such a procedure, it is
an ugly feature of the SM, often referred to as the fine-tuning problem.
There are two broad ways out of this conundrum. Either
\begin{itemize}
\item the symmetry breaking interactions become strong at a scale $\Lambda
\alt 1$ TeV, and that
form factors cut off the loop integrals at the scale $\Lambda$,
or

\item new elementary degrees of freedom not included in the SM manifest
themselves at a scale $\Lambda \alt 1$ TeV, thereby vitiating the SM
estimate of $\delta m_H^2$.
In this case, the EWSB sector
could be weakly coupled.
\end{itemize}

Assuming that we do not accept the possibility of fine-tuning, we
infer that an exploration of
the TeV energy scale {\it must} reveal new physics in some form. Our
arguments do not fix what this new physics might be. If the EWSB sector
is strongly coupled, the new physics may reveal itself as resonance states
of new elementary quanta. Technicolour, to be discussed in the next section,
is an illustration of such a scenario. The only known\footnote{Of course
the possibility that some other mechanism will be discovered in the future
remains open.} realization of a
weakly coupled EWSB sector without fine-tuning problems
is weak scale supersymmetry (SUSY), discussed
in Sec. 3. In this case, contributions to $\delta m_H^2$ from sparticles
in loops cancel the offending SM contributions, thereby alleviating
the fine-tuning issue.

There are many other extensions of the SM that have been considered in the
literature. These include new quarks or leptons (sequential or otherwise),
new gauge bosons, lepto-quarks,
coloured exotics and doubly-charged scalars, to name a few. While there is
no reason why these should not occur in nature, their existence does not
shed light on any pressing theoretical issue. Also, unlike the case of
Technicolour or low energy SUSY just discussed, the mass scale for these
is essentially arbitrary. We will refer to such exotica as Optional
New Physics. For reasons of space and time, we will not discuss
these in any detail
but will only touch on experimental constraints on these in Section 4.

\section{Strongly Coupled Electroweak Symmetry Breaking}

We first consider the case where the new physics implied by
our analysis of the symmetry breaking sector of the SM consists
of new strong interactions between the quanta of this sector.
These are expected to manifest themselves as strong interactions
in $V_LV_L$ scattering at high energy, since $V_L$, the longitudinal component
of $W$ and $Z$ bosons is dominantly composed of quanta of the EWSB sector.
These new strong interactions, if we are lucky, may also lead to formation
of new resonance states which may be accessible at future colliders.

No one has been able to come up with a convincing
and phenomenologically acceptable model in which electroweak
symmetry is dynamically broken. One
practical problem with all such
scenarios is that the strong interactions
make it difficult to perform dyamical calculations. Quite aside from
this, the nature of the strong interactions that drive EWSB
is unknown, except that they must
respect a custodial symmetry in order to yield
$\rho = \frac{M_W}{M_Z\cos\theta} \simeq 1$.

A conceptually beautiful idea goes under the name of Technicolour\cite{TECH}.
It is hypothesized that there are new (weak iso-doublet)
fermions (dubbed Techni-fermions) that
interact with one another via chirally symmmetric
Technicolour gauge interactions, in much
the same way that ordinary quarks interact via QCD. Just as chiral symmetry
between quarks
is broken at a scale where QCD interactions become strong, it is assumed
that the Technicolour interactions cause a condensate of Techni-fermions,
but with a scale  $\sim 3000\Lambda_{QCD}$, resulting in a (Techni-)
chiral symmetry breakdown and the corresponding Goldstone boson, analogous
to the pion in QCD.\footnote{Other resonances, with quantum numbers of the
observed particles, but masses generically
in the TeV range will also be present. A
problem is that often complicated Technicolour models admit
additional pseudo-Goldstone bosons with masses $O(10 \ GeV)$, and hence,
are excluded by colliders searches.}
The self energy function of the electroweak vector bosons thus develops
a dynamical pole at $k^2 = 0$, thereby giving non-zero masses to the
$W$ and $Z$ bosons\cite{SCHW}, together with the correct value of $\rho$.

Since ordinary fermions do not couple to Technicolour, their chiral symmetry
prevents them from acquiring masses along with the vector bosons. In order
to give masses to these, new extended Technicolour
(ETC) gauge interactions that
couple ordinary fermions ($f$) to Techni-fermions ($F$)
need to be postulated\cite{ETC}. The ETC
gauge bosons convey the news of chiral symmetry breaking in the
Techni-fermion sector to the ordinary fermions,
which then develop masses as shown in Fig. 1.
\begin{figure}[htb]
\vspace{45mm}
\caption{A diagramatic representation of how ETC bosons ($V_{ETC}$) convey the
information about chiral
symmetry breakdown in the Techni-fermion ($F$) sector to the ordinary
fermions ($f$), thereby inducing a mass ($m_f$) for these. The cross
denotes the Technicolour condensate which is the order parameter for
the chiral symmetry in the Techni-fermion sector.}
\end{figure}
We find,
\begin{equation}
m_f \sim \frac{g_{ETC}^2}{M_{ETC}^2} \times <...>,
\end{equation}
where $g_{ETC}$ is the ETC gauge coupling, $M_{ETC}$ the mass of the ETC
gauge boson, and $<...>$ is the {\it dynamics-dependent} Technicolour
condensate which is a measure of chiral symmetry breaking in the
Techni-fermion sector. The ETC interactions must
distinguish between
ordinary fermion flavours in order to account for the wide range of
quark masses observed in nature. This is dangerous\cite{TECHREV}
because these ETC
interactions can then induce interactions of the form,
\begin{displaymath}
\frac{g_{ETC}^2}{M_{ETC}^2}(\bar{s}\gamma^{\mu}d)^2,
\end{displaymath}
which are strongly constrained by the $K_L$-$K_S$ mass difference. In the
simplest case where Technicolour dynamics is assumed to be identical (except
for scale) to QCD dynamics, a value of $\frac{g_{ETC}}{M_{ETC}}$ large
enough to produce masses of $O$(10-100 MeV) for down and strange quarks
is incompatible with FCNC constraints.
The second strike against these models, is
that they give\cite{PESKIN} (assuming QCD-like Technicolour dynamics)
$S \simeq +1$ which is excluded, at the $4\sigma$ level, by
the data. It is also unclear whether it is
possible to accommodate a large top-bottom mass splitting without
inducing unacceptably large corrections to the $\rho$ parameter.

Walking Technicolour\cite{HOLDOM} models were invented to alleviate
the incompatibility between quark masses and FCNC constraints. In these
models, the dynamics is arranged so that the renormalized condensate in
Eq. (1) is enhanced relative to its naive value, thereby allowing for larger
fermion masses for a fixed value of $\frac{g_{ETC}}{M_{ETC}}$. It has been
argued\cite{APPEL} that these models may be able to accommodate smaller (or
even negative) values of the Peskin-Takeuchi $S$-parameter. The real problem
with all Technicolour models probably comes from the measured value
of the $\bar{b}b$ branching fraction of the $Z$. The point is that the
{\it same} ETC gauge boson that is responsible for the $t$-quark mass also
contributes a correction\cite{CHIV},
\begin{equation}
\frac{\delta\Gamma(\bar{b}b)}{\Gamma(\bar{b}b)} \simeq -3.7\%\frac{m_t}{100
GeV}\xi^2,
\end{equation}
where $\xi$ is a Clebsch of order unity.
We see that this correction tends to {\it reduce} $R_b$ from its SM value
by about 5\% if the top quark is heavy. Numerical estimates\cite{GATES}
show that while the magnitude of these corrections can be made smaller
in Walking Technicolour scenarios, they are generally large enough to
impact on the LEP measurements.
Since, as shown in Table 1,
the measured value of $R_b$ is already higher than its SM expectation, this
measurement, if it holds up, could create a serious problem for such
scenarios.

We thus conclude that the Technicolour idea, where ETC interactions
are responsible for fermion masses seems to be strongly
disfavoured by experiment.
It could, of course, be that while Technicolour
is indeed responsible for EWSB and gauge boson masses, the origin of fermion
masses lies elsewhere. Nevertheless, it is fair to say that no
phenomenologically
viable model of dynamical EWSB has as yet emerged. On the other hand,
because detailed dynamical consequences of strongly coupled theories
are difficult to compute with present day techniques, it is
impossible to absolutely exclude the Technicolour paradigm. Finally,
it could, of course, be that there is indeed a strongly coupled EWSB sector,
but that the specific realization of this provided by Technicolour is
incorrect.

These inherent uncertainties have led many authors to
adapt methods of chiral dynamics
to explore the implications of a strongly coupled EWSB sector.
It is conservatively assumed that the effects of any new physics,
whether it is Technicolour resonances, a heavy scalar resonance,
or something else, cannot
be {\it directly} explored at energies that will be experimentally accessible,
but may be indirectly
observable in the scattering of longitudinal $W$ and $Z$ bosons.
The starting
point of this approach
is the (non-renormalizable) momentum expansion of the effective Lagrangian
that describes Goldstone boson scattering, the first term of which
yields a model-independent contribution to the Goldstone boson scattering
amplitude in agreement with low energy theorems. This amplitude is
unitary only up to a scale $\Lambda$ at which effects of other physics
become significant; these effects are
embodied in the (model-dependent) non-renormalizable terms in the
effective Lagrangian.
For values of $s$ such that $M_W^2 << s \alt \Lambda$,
the equivalence theorem then tells us that the  model-dependent,
unitarized\footnote{In some models, amplitudes are really unitary. In other
models, unitarity is only assured up to an energy scale of a few TeV, which is
fine for practical purposes.}
Goldstone boson amplitudes are a good approximation
to longitudinal vector boson scattering amplitudes via which
effects of a strongly coupled EWSB may be indirectly searched for.

How does one study vector boson scattering? In high energy hadron
($e^+e^-$) collisions,
$W$ and $Z$ bosons can be radiated from the initial state quarks
(leptons) in much the same way that these radiate photons. It is
the elastic scattering of the vector bosons that is used to probe
the EWSB sector. Bagger {\it et. al.}\cite{BAGGER} have studied the
signal from $V_LV_L$ ($V = W,Z$) scattering at hadron supercolliders.
They have focussed on the signal where the vector bosons decay leptonically
to eliminate backgrounds from QCD jets. Then, the main backgrounds come
from $V_TV_T$ and $V_LV_T$ production, which has the same kinematical
characteristics as the signal, and in the case of $W_L^+W_L^-$ scattering,
also from $\bar{t}t$ production. As one might imagine, very stringent cuts
discussed in Ref.\cite{BAGGER} are needed to sort out the signal from
background. It is not our purpose here to address the details of these
computations. We will content ourselves by presenting sample results
together with a few remarks.

A sample of some results
obtained in Ref.\cite{BAGGER} are illustrated in Table 2,
where the background and signal rates at a 16 TeV $pp$ collider
in various $VV$ scattering channels
is shown after experimental cuts. The various columns refer to different
unitarization schemes, so that the variation of the signal in the
last five columns is indicative of theoretical uncertainties.
We should mention that additional unitarization schemes
(not shown here for brevity) have been considered in Ref.\cite{BAGGER} to
which we refer the reader for further details. Yet another scheme has been
discussed in Ref.\cite{HI}.
The unitarization models
considered here are, the SM with $m_H = 1$ TeV, a model with an $s$-channel
scalar isoscalar resonance with a mass of 1 TeV and width 350 GeV, an
$s$-channel vector-isovector resonance (like the Techni-rho) with M = 2 TeV
and $\Gamma = 700$ GeV, and finally, two non-resonant models where the
amplitudes obtained from the low energy theorems are unitarized by
cutting them off at the unitarity bound following Ref.\cite{CG}(LET CG), or
using the K-matrix method.
\begin{table}[htb]
\centering
\caption[]{\small Signal event rates per 100 $fb^{-1}$ at a 16 TeV $pp$
collider for $m_t = 140$ GeV, with cuts given in Ref.\cite{BAGGER} for
several models discussed in the text. The Table is adapted from
Ref.\cite{BAGGER}}
\bigskip
\begin{tabular}{|l|c|c|c|c|c|c|} \hline
Channel & Bkgd & SM & Scalar & Vec 2.0 & LET CG & LET K \\ \hline
%
%
$ZZ$ & & & & & &  \\
$M_{ZZ}>0.5$ & 1.0  & 14  & 7.5 & 1.4 & 2.5 & 2.2\\
$M_{ZZ}>1.0$ &0.1 & 3.9  & 2.7 & 0.4 & 1.1 & 0.9\\ \hline
$W^+W^-$ &  &  &  &  &  &  \\
$M_{\ell\ell}>0.25$ & 18 & 40 & 26 & 8 & 9.2 & 7.2 \\
$M_{\ell\ell}>0.5$ & 15 & 32 & 21 & 7.4 & 8.3 & 6.3 \\ \hline
$W^+Z$ &  &  &  &  &  &  \\
$M_T> 0.5$ &2.4 & 1.0 & 1.4 & 4.8 & 3.2 & 2.9 \\
$M_T> 1.0$ &0.3 & 0.3 & 0.4 & 3.3 & 1.6 & 1.4 \\ \hline
$W^+W^+$ &  &  &  &  &  &  \\
$M_{\ell\ell} >0.25$ & 6.2 & 9.6 & 12 & 12 & 27 & 24 \\
$M_{\ell\ell} >0.5$ & 1.7 & 3.7 & 5.2 & 4.8 & 16 & 14 \\ \hline
\end{tabular}
\end{table}
It has been claimed that there is a statistically significant
signal which should be visible
above the background in at least one of the channels after two years
of LHC running with a luminosity of 100 $fb^{-1}$
for all the
models considered. We note, however, that the event rate is frequently
very small --- a few events per 100 $fb^{-1}$ at the LHC. Even if it can
be argued that the physics backgrounds have been carefully analysed, the
smallness
of the signals should cause worry about reducible, detector-dependent
backgrounds. Other significant concerns might be how long it would actually
take the LHC experiments to accumulate 200 $fb^{-1}$ of data, and whether
it will be possible to effectively
implement the stringent cuts of Ref.\cite{BAGGER}
in the high luminosity LHC environment. It is
worth mentioning that a study of these signals is the one area where
the SSC had a clear advantage over the LHC. The design energy of the
LHC has been reduced from 16 TeV in Table 2 to 14 TeV. Reducing it
further would certainly make physics discovery in this channel extremely
difficult. Indeed, untangling
this signal, if nature has indeed chosen the strongly coupled EWSB sector
option with no accessible new resonance states,
will be a real challenge for our experimental colleagues.

The feasibility of detecting departures from a weakly coupled EWSB sector
by studying $V_LV_L$ scattering at linear $e^+e^-$ colliders has also been
investigated. We will not discuss this in detail but will refer the
reader to the review in Ref.\cite{HAN}. Here, the clean environment makes it
possible to use the hadronic decay modes of the $W$ and $Z$ bosons, but
has the disadvantage that $W_L^{\pm}W_L^{\pm}$ scattering amplitudes, which
could help sort out dynamical issues, are not accessible.
Since only a small fraction of energy goes into $V_LV_L$ scattering, the signal
rates are small, and frequently, comparable to background. A 1.5 TeV collider
with an integrated luminosity of $\sim$~200~$fb^{-1}$ is the minimum that
is required to be able to probe this physics. In addition, it is important
to be able to distinguish hadronically decaying $W$ and $Z$ decays in order
to be able to examine the underlying dynamics by separating $W^+W^-$ and $ZZ$
processes.

Finally, we briefly remark on the feasibility of studying strong $V_LV_L$
scattering at $\gamma\gamma$ colliders. It has been known for some time that
the signal processes $\gamma\gamma \rightarrow W_LW_L, Z_LZ_L$ are swamped by
backgrounds\cite{BACK}
from $\gamma\gamma \rightarrow W_TW_T, Z_TZ_T$ reactions. Very recently,
it was pointed out\cite{BRODSKY} that in high energy collisions,
photons may radiate
$W_L$'s in the same way
that they are radiated from electrons and
positrons at Linear Colliders, except, of course, that the $W$ radiation
from photons must occur in pairs. The idea then is to use the luminosity
of $W_L$'s in resolved photon collisions to study the
$W_L^{\pm}W_L^{\pm} \rightarrow W_L^{\pm}W_L^{\pm}$,
$W_L^+W_L^- \rightarrow W_L^+W_L^-, Z_LZ_L$ scattering amplitudes by tagging
the two spectator $W$ bosons to eliminate backgrounds from
$\gamma\gamma \rightarrow W_TW_T, Z_TZ_T$ processes. With an integrated
luminosity of 100 $fb^{-1}$ and an overall efficency of about 10\%, a
preliminary study\cite{CHEUNG} suggests
that a signal should be observable in 1.5 TeV
$\gamma\gamma$ collisions. Somewhat less optimistic conclusions have been
obtained by Jikia\cite{JIK} who has incorporated a realistic photon spectrum
into the analysis.

\section{Weakly Coupled Electroweak Symmetry Breaking: Low Energy SUSY}
\subsection{Motivation and Framework}
If there are no strong interactions in the EWSB sector, then the unwanted
large SM corrections to the Higgs boson mass discussed in Sec. 1 must be
cancelled by other
corrections from new physics that must enter by the TeV
scale. This cancellation, between SM loop contributions and additive
new physics
contributions need not be exact, but must be good to many significant figures
in order that two contributions, each of $O(\Lambda^2)$ (recall that $\Lambda$
could have been as large as $M_{GUT}$ or even $M_{Planck}$
without the intervention
of new TeV scale physics), conspire to yield a result $\sim$ (100 GeV)$^2$.
It is our prejudice that such a precise cancellation cannot be accidental,
but must result from
a symmetry between SM physics and the new physics in question.
Since bosonic and fermionic loops contribute with opposite signs to
$\delta m_H^2$, such a cancellation is possible in a theory
that contains bosons and fermions, with masses and couplings to the Higgs boson
related in a way that provides this cancellation. Notice that a symmetry that
accomplishes this differs from all known symmetries ({\it e.g.}
gauge symmetries, Lorentz invariance, {\it etc.}) in that it relates properties
of bosons and fermions. Such a symmetry is called a supersymmetry.

Low energy supersymmetry\cite{SUSYREV} provides the only known way of
incorporating elementary scalar bosons into the SM, without the need for
fine-tuning order by order in perturbation theory. The idea is that for each
``quadratically divergent'' contribution from a SM boson (fermion) loop,
there is a corresponding contribution from a loop with
a new fermion (boson), the
supersymmetric partner of the SM particle\footnote{Since the particles and
their superpartners have the same gauge quantum numbers, it is not possible
to identify SM bosons and fermions as superpartners of one another.} which
cancels against it. One may wonder how this comes about when the loop
integrals depend on the masses and the couplings of the various particles.
The point is that supersymmetry implies the particles and their
superpartners have exactly the same masses and
couplings (aside from Clebsch-Gordan factors). This is, of course,
unacceptable on phenomenological grounds since the SUSY partners of at
least the charged particles would have long since been discovered.
Fortunately, these cancellations need not be exact. Since it suffices
to arrange matters so that
\begin{displaymath}
\delta m_H^2 \alt (1 \ TeV)^2,
\end{displaymath}
we need only to require,
\begin{equation}
m_{sparticle}^2 - m_{particle}^2 \alt (1 \ TeV)^2.
\end{equation}
Eq. (3) provides the rationale for TeV scale supersymmetry. Within this
framework, a weakly coupled
EWSB sector can be stable to radiative corrections provided only that
sparticle masses are smaller than about 1~TeV. These sparticles
may be then be detectable
at the next generation of high energy colliders.

The simplest supersymmetric extension\cite{SUSYREV} of the SM, known
as the Minimal
Supersymmetric Model (MSSM), may be obtained by a direct
supersymmetrization of the SM. For every chiral fermion ($f_i, i=L,R$) in the
SM, there is a spin zero particle $\tf_i$ (the squarks and sleptons,
collectively referred to as sfermions) with the same colour and electroweak
quantum numbers. Any
$\tf_L$-$\tf_R$ mixing is proportional to the corresponding {\it fermion}
mass, and hence, is negligible except for top squarks. We will also ignore
any flavour mixing between squarks and,
therefore, assume that the the sleptons ($\tell_i$) and squarks
($\tq_i$, $q \neq t$) are also mass eigenstates. Notice that gauge
symmetry completely fixes the gauge interactions of these sfermions.
Turning to the gauge sector, spin-$\frac{1}{2}$ gauginos, in the adjoint
representation of the gauge group, are the SUSY
partners of the gauge bosons. These include the colour octet gluino ($\tg$),
the charged winos, and the neutral photino and the zino. Because it is not
possible to simultaneously couple fermions to both
the Higgs field and its complex
conjugate in a manner consistent with supersymmetry, even the simplest
SUSY extension requires two doublets that acquire vacuum
expectation values (vevs) in order to give masses to both
up and down type quarks. After the Higgs mechanism, two
neutral scalars $H_\ell$ and $H_h$, a pseusoscalar $H_p$ and a pair
of charged scalars $H^{\pm}$ are left over in the physical spectrum.
The spin $\frac{1}{2}$ SUSY partners of the two Higgs boson doublets,
the Higgsinos, mix with the gauginos of the same charge once electroweak
symmetry is broken. The physical particles are two Dirac charginos ($\tw_1$
and $\tw_2$) and four Majorana neutralinos ($\tz_i$, $i=1...4$) labelled
in order of increasing mass, along with gluinos which, being colour octets,
cannot mix with anything else.

As we have seen, SUSY must be a broken symmetry. For phenomenological
purposes, it is sufficient to parametrize SUSY-breaking by introducing
all terms consistent with gauge and space-time symmetries that do not
lead to the reappearance of quadratic divergences we have worked so hard
to eliminate. All such terms, known as soft SUSY breaking terms, have been
classified in Ref.\cite{GIRARDELLO}. They consist of sfermion
(and also Higgs scalar) and gaugino (but not matter fermion or Higgsino)
masses, along with the trilinear
and bilinear SUSY breaking scalar couplings, which are generally
unimportant for phenomenology except in the scalar top sector.

Without further assumptions, the number of soft-SUSY breaking parameters is
very large, making phenomenological analyses intractable. For instance, each
SU(3)$\times$SU(2)$\times$U(1) scalar multiplet has an independent
SUSY-breaking mass, and also,
the three gaugino masses are {\it a priori} unrelated.
It is generally assumed that the gaugino masses all arise from a common
GUT gaugino mass, and so, unify at some ultra-high scale. Within the
GUT framework, scalars within a common GUT multiplet, of course, have
the same mass\footnote{For example, in SU(5) there would be two mass
parameters for each generation but just one mass per generation in
SO(10)}, since SUSY is assumed to be broken only at the TeV scale. This
mass is defined at the unification scale and, of course, must be
evolved\cite{RGE} down to the electroweak scale relevant for phenomenology.
This leads to a splitting between the physical sparticle
masses\footnote{This splitting has often been ignored in many early MSSM
analyses where squarks and sleptons are all taken to be degenerate.}.
But even in
a GUT, there are still a large number of free mass parameters. Motivated
by supergravity models (which we will discuss in more detail later), it is
frequently assumed that there is a single mass gap in the scalar matter
sector, so that all the squarks and sleptons have a common mass at the
unification scale. Evolution of these masses down to the weak scale
removes this degeneracy. The dominant splitting comes from QCD interactions,
resulting in a squark-slepton mass difference given by,
\begin{equation}
\overline{m_{\tq}^2} - \overline{m_{\tell}^2} \simeq 0.77m_{\tg}^2,
\end{equation}
where $\overline{m_{\tq}^2}$ ($\overline{m_{\tell}^2}$)
denotes the average squark (slepton) mass squared. The sfermions are
also further split by electroweak interactions and by the so-called
$D$-terms not shown here.

Finally, we mention that the gauge interactions of sparticles automatically
conserve $R$-parity, which is a multiplicative quantum number defined to be
+1 for ordinary particles, and -1 for their SUSY partners. In the MSSM
framework, it is assumed
that R-parity is conserved by {\it all} interactions. While the introduction
of additional global or discrete symmetries is not particularly desirable, the
introduction of $R$-parity conservation, or alternatively,
the conservation of baryon or
lepton number is absolutely essential to prevent proton decay at a
catastrophic rate. Important consequences of $R$-parity conservation are
that sparticles can only be produced in pairs by collisions of ordinary
particles, and that the lightest supersymmetric particle (LSP) is absolutely
stable. Cosmological arguments\cite{SUSYREV} then imply that it must be
colour and electrically neutral. The LSP is a candidate for the
dark matter content of the universe.
Of the sneutrino and neutralino candidates for the LSP, the sneutrino
is strongly disfavoured if we also assume that the LSP is also the
{\it galactic} dark matter\cite{DM}.
$R$-parity violating models which lead to very interesting
phenomenology\cite{RPARITY} can also be constructed. We will not have time to
discuss these here.

With these assumptions, the model is completely determined by
the parameters, $m_{\tq}$, $m_{\tg}$, $\tan\beta$, the ratio of the vevs
of the two Higgs fields, $\mu$, the supersymmetric Higgsino mass,
$m_{H_p}$ and $A_t$, the SUSY violating
trilinear Higgs-stop scalar coupling which
mainly affects only top squark phenomenology. We are now ready to
study signals for sparticle production in high energy collisions
and discuss how the non-observation of any signal to date serves to
constrain the model parameters.

\subsection{Constraints from Collider Experiments}

The cleanest limits on sparticles come from the experiments at LEP.
The agreement\cite{OLCH} between the measured value of $\Gamma_Z$ and its SM
expectation implies an upper limit on the decays of the $Z$ boson into
sparticles. This translates to a lower limit\cite{DM,DT} of $\sim$35-40~GeV
on the mass of the charged sparticles (and also sneutrinos if these
decay visibly) which is independent of how the sparticles decay. Bounds
from the invisible width of the $Z$ yield similar limits on $m_{\tnu}$
if these decay invisbly via $\tnu \rightarrow \nu \tz_1$. The corresponding
limits on neutralino masses are very sensitive to model parameters, since
$\tz_1$ and $\tz_2$ couplings to the $Z$ become negligible when these are
gaugino-like, as is the case for $\mu \agt m_{\tg}$. Searches in exclusive
channels\cite{LEP},
assuming that sfermions decay via $\tf_i \rightarrow f\tz_1$,
and $\tw_1 \rightarrow f\bar{f}\tz_1$ (which lead to spectacular $\eslt$
events) yield lower mass bounds on sfermion and $\tw_1$ masses that are
very close to $\frac{M_Z}{2}$. The non-observation of any signals from
the decays $Z \rightarrow \tz_1\tz_2, \tz_2\tz_2$ also excludes some
regions of the MSSM parameter space.
In fact, the clean environment of $e^+e^-$
collisions should enable the discovery of charged sparticles all the
way up to the kinematic limit, so that LEP II should be able to probe chargino,
slepton and squark masses up to about 80-90~GeV.

Experiments at hadron colliders are best suited for the detections of squarks
and gluinos. As is well known, gluinos and squarks in the mass range currently
being probed at the Tevatron decay via a complicated cascade\cite{CAS} which
terminates in the LSP. The various sparticle pair production mechanisms at
hadron colliders
together with the
cascade decay patterns of all the sparticles as given by the MSSM, have been
incorporated into ISAJET\cite{ISAJET} which can now be used to simulate SUSY
signals in the CDF and D0 experiments currently operating at the Tevatron.

The classic signature of gluinos and squarks is $\eslt$ from the
undetected LSPs produced at the end of each cascade. The non-observation of
an excess of $\eslt$ events
at the Tevatron has enabled\cite{TEVGL} the D0 collaboration to infer lower
limits around 150~GeV on their masses, improving on the earlier limit of
$\sim 100$~GeV obtained by CDF; if $m_{\tq}=m_{\tg}$, the mass bound improves
to about 205~GeV, since then both squarks and gluinos can contribute to the
signal. In this analysis, which is based on about 15 $pb^{-1}$ of data,
it is assumed that all squarks other than $\tt$
have the same mass.
The region of the $m_{\tg}-m_{\tq}$ plane excluded by the CDF and D0
analyses of their data is shown in Fig. 2.
\begin{figure}[hbt]
\vspace{115mm}
\caption{The region
of the $m_{\tq}-m_{\tg}$ plane excluded by the search
for $\eslt$ events at the Tevatron for ten degenerate squark flavours. This
region is weakly dependent on $\mu$ and $\tan\beta$ which have been fixed at
-250~GeV and 2, respectively. This figure is from the analysis by the D0
Collaboration[44].}
\end{figure}
The Tevatron experiments are each
expected to accumulate 50-100 $pb^{-1}$ of integrated luminosity by the end
of the current run. This should enable them to probe gluino masses up to
about 250-300~GeV, both via $\eslt$ searches and by way of multi-lepton signals
discussed below.

Before proceeding further, it is fair to ask what the measured
value of $\Gamma(b\bar{b})$, which caused so much problem for Technicolour
models, implies in the SUSY framework. We note that SUSY, being a decoupling
theory, cannot fair worse than the SM. The question, therefore, is whether
sparticle loops can increase the prediction of $R_b$, bringing it closer
to its measured value. This has recently been examined in Ref.\cite{WELLS}.
Assuming that the top quark mass is within 1$\sigma$ of the central
value obtained by CDF\cite{TOP}, these authors find that within the MSSM
framework, SUSY contributions from chargino-top squark loops can bring
$R_b$ within 1$\sigma$ of its measured value if at least one of the sparticles
is lighter than 65~GeV.\footnote{However, within the constrained framework
that they advocate, they conclude that the 2$\sigma$ discrepancy cannot
be significantly reduced.}
We should view this in proper perspective. It is not too important
at the present time to reduce the discrepancy to below $1\sigma$. What is
important, of course, is that SUSY theories do {\it not} necessarily
lead to a further reduction of $R_b$ as in the case of Technicolour.

\subsection{SUSY Search in the 1990's}

The Fermilab Tevatron and HERA are expected to remain the highest
energy $\bar{p}p$ and $ep$ colliding facilities till well beyond
the turn of the millenium, while LEP II, which should begin operation
in just about a year, will be the energy frontier for $e^+e^-$ collisions.
As already noted, the cleanliness of $e^+e^-$ collisions will enable
experiments at LEP II to cleanly probe charged sparticle masses up to
80-90~GeV. LEP II will also be able to search for Higgs bosons with masses
up to about 90~GeV. As we have discussed, this is a substantial portion of
the allowed parameter space, if we believe that Higgs boson interactions
remain perturbative up to a very large energy scale. This is especially the
case for the Higgs bosons in supersymmetry, as has been emphasized in
Ref.\cite{SUSYHIG}.

At HERA, the most important SUSY processes are $q + e \rightarrow \tq + \te$
and $q + e \rightarrow \tq + \tnu$. Within the MSSM,
Cashmore {\it et. al.}\cite{CASHMORE}
have concluded that the range of sparticle masses that
can be explored with an integrated luminosity of 200 $pb^{-1}$ at HERA,
is roughly given by $m_{\tq} + m_{\te} \alt 180$~GeV, $m_{\tq} + m_{\tnu}
\alt 130$~GeV.
The mass reach for other sparticles is even smaller.
In view of the bounds on sparticle masses that have
already been attained in experiments at the Tevatron and LEP, it appears
extremely unlikely that HERA will be a discovery machine for supersymmetry.
The situation may, however, be quite different if $R$-parity is not conserved.
If $R$-parity is violated by electron lepton number violating interactions,
it is quite possible that squarks can be singly produced (as resonances
in $eq$ scattering) at HERA. We refer the reader to Ref.\cite{DREINER} for
a study of the resulting signals.

The Tevatron experiments will collectively accumulate an integrated
luminosity in excess of $\sim$ 100 $pb^{-1}$ by the end of the current run,
and are expected to improve on this by an order of magnitude after the Main
Injector begins operations. In addition to extending the $\eslt$ search
region for gluinos and squarks, the
large increase in the data sample should make it possible to search for
supersymmetry in many other channels. The most promising of these are,
({\it i})~gluino and squark searches
via multilepton events from their cascade
decays, ({\it ii})~search for $\tw_1\tz_2$ production via isolated
trilepton events free of jet activity, and ({\it iii})~search for the
lighter $t$-squark, $\tt_1$. We note that Tevatron experiments will not be
able to probe slepton masses significantly beyond the reach of LEP.

{\it Multilepton Signals from Gluinos and Squarks.} The conventional $\eslt$
search for $\tg$ and $\tq$ is background limited. Even with an integrated
luminosity of 1 $fb^{-1}$ that should be available with the Main Injector
upgrade of the Tevatron, we anticipate a maximum reach of $\sim 270$~GeV
($\sim 350$~GeV) if $m_{\tq} >> m_{\tg}$ ($m_{\tq} \simeq m_{\tg}$) in this
channel\cite{KAMON}.
Heavy gluinos and squarks can also decay via the chargino and
$\tz_2$ modes which, unless suppressed by phase space,
frequently dominate the decays of $\tq_L$ and $\tg$. The subsequent leptonic
decays of the $\tw_1$ and $\tz_2$ yield events with hard jets accompanied by
1-3 isolated, hard leptons and $\eslt$. The cross sections for various
multilepton topologies, after cuts\cite{BKTGL}
to simulate experimental conditions at
the Tevatron, are shown in Fig.~3 for different choices of gluino and
squark masses.
\begin{figure}[hbt]
\vspace{90mm}
\caption{Total cross sections for $\eslt, 1\ell,2\ell,3\ell,4\ell$ and
same-sign (SS) dilepton event topologies in 1.8 TeV $p\bar{p}$ collisions.
We have fixed $\mu = -200$~GeV, $\tan\beta =2$, $m_t = 140$~GeV and $m_{H_p}
= 500$~GeV. The experimental cuts as well as physics background levels are
described in detail in Ref.[50] from which this figure is taken.}
\end{figure}

While there are substantial
backgrounds to $1\ell$ and $\ell^+\ell^-$ event topologies, the {\it physics}
backgrounds in the $\ell^{\pm}\ell^{\pm}$ and $3\ell$ channels mainly
come from $t\bar{t}$ production, where a secondary lepton from the decay
of the daughter $b$ quark, which is usually inside a jet,
is accidently isolated. These backgrounds have been evaluated
in Ref.\cite{BKTGL} and
are essentially
negligible, especially if $m_t > 160$~GeV; assuming detector dependent
non-physics backgrounds can be controlled,
the 3 $\ell$ and
$\ell^{\pm}\ell^{\pm}$ search channels are essentially rate limited. These
channels, which at the Main Injector have a reach\cite{BKTGL} of 230-300~GeV
depending
on $m_{\tq}$ and $m_{\tg}$, provide complementary ways of searching for
gluinos and squarks at the Tevatron, and because they are free of SM
backgrounds, may even prove superior to conventional $\eslt$ searhces
if gluinos and squarks are very heavy.
We remark that detection of these signals provides (indirect) evidence
for charginos and neutralinos beyond the range of LEP.

{\it Search for Isolated Trilepton Events.} Associated $\tw_1\tz_2$ production
which occurs by $s$-channel $W^*$ and $t$-channel $\tq$ exchanges, followed by
the leptonic decays of $\tw_1$ and $\tz_2$ results in isolated trilepton plus
$\eslt$ events, with hadronic activity only from QCD radiation. If $\tw_1$
and $\tz_2$ are light enough, $\tw_1\tz_2$ production is dominated by
the decay of on-shell $W$-bosons, so that the trilepton cross section is
very large\cite{RESW}. However, the region of parameter space where
this is possible is now already excluded by the LEP bound on $m_{\tw_1}$.
Nath and Arnowitt\cite{NA} had pointed that {\it non-resonant}
$\tw_1\tz_2$ production will also lead to observable signals, once Tevatron
experiments accumulate an integrated luminosity $\sim$~100~$pb^{-1}$. It was
subsequently noted\cite{BT} that the leptonic decays of $\tz_2$, and
sometimes also of $\tw_1$, can be considerably enhanced if sleptons are
substantially lighter than squarks and $\mu \agt m_{\tg}$, as is the case,
{\it e.g.} in
the no-scale limit of supergravity models\cite{NOSCALE}.

SM physics backgrounds to the $3\ell + n_{jet}\leq 1$ signal are negligible,
assuming that $WZ$ events can be vetoed with
high efficiency by requiring $m_{\ell\bar{\ell}}\not=M_Z$ within experimental
resolution; a conclusive observation of a handful of such events could,
therefore,
be a signal for new physics, depending on how well detectors can veto
fake backgrounds from jet-lepton misidentification\footnote{In fact, a recent
detailed study\cite{KAMON} has concluded that Drell-Yan and $Z \rightarrow
\ell\bar{\ell}$ events where an additional jet fakes a lepton is the main
background to this signal. It would be of interest to study whether this can
be eliminated, without significant loss of signal by requiring  $\eslt \agt
20$~GeV.}
Preliminary analyses by the CDF and D0
experiments\cite{WINO}
(for large values of the Higgsino mass parameter, $\mu$)
are already competitive with bounds from LEP.
The experiments will soon explore\cite{BKTWIN,ZICH,KAMON}
parameter ranges not accessible at LEP
and, under favourable circumstances, may be competitive with LEP II. Within
the MSSM framework, this reach translates to $m_{\tg} = 250-400$~GeV depending
on $m_{\tq},\mu$ and $\tan\beta$, the ratio of the two Higgs vacuum
expectation values.

{\it Searching for Top Squarks at the Tevatron.}
Third generation squarks differ from other squarks in that they have
large Yukawa interactions.
These interactions affect the mass of the squarks in two distinct ways.
First, they reduce the diagonal masses of the $\tt_R$ and the $\tt_L$ (and by
SU(2) invariance, also of $\tb_L$) squarks via their contributions to the
running of squark masses. Second, they mix $\tt_L$ and $\tt_R$
further reducing the mass of $\tt_1$, the lighter of the
two mass eigenstates. In fact, it is theoretically (but, of course, not
phenomenologically) possible that
$\tt_1$ is essentially massless with other squarks and gluinos all too heavy
to be produced at the Tevatron.
The Tevatron lower limits on $m_{\tq}$ are derived assuming ten
degenerate squark flavours, and thus, are not applicable to $m_{\tt_1}$.
Currently,
the best limit, $m_{\tt_1}\alt M_Z/2$, comes from LEP experiments; this bound
can be evaded if the stop mixing angle and $m_{\tt_1}-m_{\tz_1}$ are {\it
both} fine-tuned, a possibility we do not consider here.

The signals from top squark production depend on its decay patterns. For
$m_{\tt_1}< 125$~GeV, the range of interest at the Tevatron,
$\tt_1$ decays via the loop-mediated mode, $\tt_1\to c\tz_1$
if the
tree level decay $\tt_1\to b\tw_1$ is kinematically forbidden\cite{HK}.
Stop pair production is then signalled by $\eslt$ events from its direct
decays to the LSP, and so, can be searched for via the canonical $\eslt$
search for SUSY. A recent Monte Carlo analysis has shown that,
with a data sample of 100 $pb^{-1}$,
Tevatron experiments should be able\cite{BST}
to probe stop masses up to 80-100~GeV, even if $\tz_1$ is relatively heavy.

The tree level
mode $\tt_1 \rightarrow b\tw_1$
dominates stop decays whenever it is kinematically allowed.
The subsequent leptonic
decay of one (or both) of the charginos lead to single lepton (dilepton)
+ $b$-jet(s)+$\eslt$ events, very similar to those expected from $t\bar{t}$
pair production. Top production is thus a formidable
background to the stop signal\cite{GUN}.
For $m_t=175$~GeV, $m_{\tt_1} = 100$~GeV
and $m_{\tw_1}=70$~GeV, we have estimated that
stop events would contribute about
33\% (20\%) of the recently published CDF
sample of top candidates\cite{TOP} in the $1\ell$ (dilepton) channel.
Thus $t$-squark production could be the culprit
if Tevatron
experiments conclusively measure a top cross section significantly above SM
expectation.\footnote{Even though the central
value of $\sigma_{t\bar{t}}$ is larger than the SM expectation for
$m_t=174$~GeV (the CDF central value of mass measured from event
characteristics), there is no statistically significant discrepancy at
present.} Special cuts need to be devised to
separate the stop signal
from top events. Since
stops accessible at the Tevatron are considerably lighter than
$m_t$, and
because the chargino, unlike $W$, decays via three body modes into
a {\it massive} LSP, stop events are generally softer than top events.
It has
been shown\cite{BST} that by requiring $m_T$($\ell\eslt$)$<45$~GeV, and
$n_{jet} \leq 4$ in addition to other canonical
cuts, stops with masses up to about 100~GeV should be detectable
in the $1\ell$ channel by the
end of the current Tevatron run, assuming a $B$-tagging efficiency of
30\% for $B$ hadrons with $p_T > 15$~GeV and $|\eta_B| \leq 1$.
In the dilepton channel, the stop events can be selectively enhanced over
those from top production by
requiring $p_T(\ell^+)+p_T(\ell^-)+\eslt<100$~GeV.
With an integrated luminosity of
100 $pb^{-1}$ Tevatron experiments should be able to detect stops
up to about 80-100~GeV without the need for any $B$-tagging capability.

\subsection{Supersymmetry Searches in the Next Millenium}

Direct searches at the Main Injector and LEP II will probe
sparticle masses between 80-300~GeV; even assuming MSSM mass patterns, the
chargino search, by inference, will
probe gluino masses up to about 400~GeV. Since the SUSY mass scale
could be as high as $\sim$ 1~TeV, it will, unless sparticles have already been
discovered, be up to
supercolliders such as the LHC at CERN or an
$e^+e^-$ Linear Collider (*LC) to explore the remainder of the parameter
space. There has also been some talk about possible upgrades of the Tevatron
beyond the Main Injector. We will touch upon these only very briefly in
our discussion.

In the $\eslt$ channel, the LHC can search\cite{LHC,ATLAS}
for gluinos and squarks
with
masses between 300~GeV to 1.3-2~TeV, depending on $m_{\tq}$ and $m_{\tg}$.
It is instructive to note that
several multilepton signals {\it must} simultaneously be present\cite{BTW}
if any
$\eslt$ signal is to be attributed to squark and gluino production, though
the various
relative rates could be sensitive to the entire sparticle spectrum.
The rate for like-sign dilepton plus $\eslt$ events is enormous
for $m_{\tg} \leq 300$~GeV; this ensures there is no window between the
Tevatron and the LHC where gluino of the MSSM may escape detection\cite{BTW}.
With 10 $fb^{-1}$ of luminosity, it should be possible to search for gluinos
up to about 1~TeV in this channel\cite{LHC,ATLAS}.
Gluinos
and squarks may also be a source of high $p_T$ $Z + \eslt$ events at the
LHC, but this signal is very sensitive to model parameters.
The LHC can also search for ``hadron-free'' trilepton events from
$\tw_1\tz_2$ production. Backgrounds from top quark production can be very
effectively suppressed\cite{BCPT} by requiring that the two hardest
leptons have the
same sign of charge. The signal becomes unobservable when the two-body
decays $\tz_2\to (Z$ or $H_{\ell}) + \tz_1$ become accessible. The dilepton
mass distribution in $\ell^+\ell^-\ell'$ events can be used to reliably
measure $m_{\tz_2}-m_{\tz_1}$.
Selectrons and smuons with masses up to 250~GeV (300~GeV
if it is possible to veto central jets with an efficiency of 99\%) should
also be detectable\cite{SLEP}. Finally, we note that
with an integrated luminosity
of 100 $fb^{-1}$ the $\gamma\gamma$ decays of scalar stoponium has been
argued to allow for the detection of $\tt_1$ with a mass up to 250~GeV,
assuming $\tt_1\to b\tw_1$ (and, perhaps, also $\tt_1 \to bW\tz_1$)
is kinematically forbidden\cite{DN}.

The SUSY reach of various hadron colliders is summarized in Table 3.
In addition to the Main Injector and the LHC, we have also shown
the reach for two possible upgrades of the Tevatron that have been much
talked about recently. The results for these have been abstracted from
Ref.\cite{KAMON} to which we refer the reader for details.

\begin{table}[htb]
\centering
\caption[]{\small Discovery reach of various options of future hadron
colliders.
The numbers are subject to $\pm 15\%$ ambiguity. Also, the clean isolated
trilepton signals are sensitive to other model parameters; we show
representative ranges from the supergravity analysis of
Ref. \cite{KAMON}, where $|\mu|$ is constrained
to be large. If $|\mu| < 100-150$ GeV,
the reach may be even smaller\cite{BKTWIN}.}
\bigskip
\begin{tabular}{|l|r|r|r|r|r|}\hline
&Tevatron &Main Injector&Tevatron$^*$&DiTevatron&LHC\\
Signal& 0.1~fb$^{-1}$&1~fb$^{-1}$&10~fb$^{-1}$&
1~fb$^{-1}$&10~fb$^{-1}$\\
&1.8~TeV&2~TeV&2~TeV&4~TeV&14~TeV\\
\hline
5$\eslt (\tq \gg \tg)$ & $\tg(210)$ & $\tg(270)$ &
$\tg(340)$ & $\tg(450)$ & $\tg(1300)$ \\
$\eslt (\tq \sim \tg)$ & $\tg(300)$ & $\tg(350)$ &
$\tg(400)$ & $\tg(580)$ & $\tg(2000)$ \\
$l^\pm l^\pm (\tq \sim \tg)$ & $\tg(170)$ & $\tg(250)$ &
$\tg(275)$ & $\tg(400)$ & $\tg(1000)$ \\
$\tg ,\tq\rightarrow 3l$ $(\tq \sim \tg)$ & $\tg(200)$ &
$\tg(270)$ & $\tg(300)$ & $\tg(500$--$550)$ & $\tg(\stackrel{>}{\sim}1000)$ \\
$\tw_1 \tz_2 \rightarrow 3l$ & $\tg(200)$ &
$\tg(250-450)$ &  $\tg(300-500)$ & $\tg(250-470)$ & $\tg(400$--$700)$ \\
$\tilde{t}_1 \rightarrow c \tz_1$ & $\tilde{t}_1(80$--$100)$ &
$\tilde{t}_1 (120)$ & & &\\
$\tilde{t}_1 \rightarrow b \tw_1$ & $\tilde{t}_1(80$--$100)$ &
$\tilde{t}_1 (120)$ & & &\\
$\tl \tl^*$ & $\tl(45-50)$ & $\tl(50)$ & $\tl(50)$ & &
$\tl(250$--$300)$\\
\hline
\end{tabular}
\end{table}

We see that the Tevatron upgrades (especially the DiTevatron) mainly
lead to a significant improvement in the reach for gluinos and squarks.
The improvement over the Main Injector in the reach
for the trilepton signal from $\tw_1\tz_2$ production appears marginal,
at best. This conclusion may be altered if it turns out to be possible to
eliminate the reducible backgrounds (see Ref.\cite{KAMON})
as discussed earlier. We also note
that it is only at the LHC that the full range of sparticle masses, allowed
by our general considerations of EWSB, can be explored.

Charged sparticles (and sneutrinos, if $m_{\tnu} > m_{\tw_1}$)
should be readily detectable at an $e^+e^-$ Linear Collider, essentially
all the way up to the kinematic limit.
Unfortunately, most detailed $e^+e^-$
studies to date do not incorporate cascade decays of
sparticles.\footnote{$e^+e^-$ to sparticle pair reactions are included in
ISAJET 7.11.} This is, of
course, irrelevant for the production of the lightest of the charged
sparticles, but could be important for the signals from the more massive ones.
For SUSY models where the gaugino masses arise from a single unified gaugino
mass so that $m_{\tw_1} \sim (0.3-0.4)m_{\tg}$, a linear collider with a
reach of about 500~GeV for charginos would have the same discovery potential
as the LHC.

The real
power of these machines, however, lies in the ability to
do precision experiments which can then be used to measure sparticle masses
with an accuracy of 1-3\%\cite{MUR}, along with other MSSM parameters.
These measurements can then be used to test
various model assumptions. For instance, the analysis of Ref.\cite{FENG}
illustrates how
a study of $\tw_1$-pair production even at LEP II can, in the favourable case
of gaugino-like charginos, constrain the values of {\it certain} combinations
of MSSM parameters. These authors find that
if chargino pair production is
kinematically accessible, measurements at LEP II could
significantly constrain the ratio of the SU(2) and hypercharge gaugino
masses, and thus directly test the gaugino mass unification hypothesis.
The availability of beam polarization at future Linear Collider
facilities is a definite advantage of these machines over hadron
supercolliders. Polarized beams are useful for two reasons. First, they
obviously increase the number of observables.
Second, they can be used to reduce SM backgrounds; for instance $W$-pair
production, which is a major background to the acollinear dilepton signal
from slepton pair production, can be virtually eliminated\cite{MUR}
by using a beam of right-handed electrons.
Precision measurements of masses and SUSY parameters
will allow\cite{MUR} direct tests of
supergravity GUT
models discussed in the next subsection.
A related, and perhaps, even more basic, issue is whether such measurements
allow us to test the supersymmetry between the couplings of any sparticles that
may be discovered --- for instance, SUSY relates
the gaugino-fermion-sfermion coupling
to the appropriate gauge coupling. Recent analyses\cite{INS,JON}
suggest that the answer to this question is affirmative, at least if the
model parameters happen to lie in favourable ranges.

Before closing this section, we remark that the *LC is
also the optimal facility to study the Higgs sector of SUSY\cite{JANOT}.
As is well known\cite{HIGREV}, even with optimistic detector assumptions,
there are regions of MSSM parameter space where {\it none} of the MSSM Higgs
bosons may be detectable either at LEP II or at the
LHC.\footnote{It has been
proposed\cite{HIGB} that with sufficient
$B$-tagging capability, it may be
possible to fill up this ``hole'' in parameter space;
whether this is realistic in the high
luminosity environment of the LHC is still unclear.} Furthermore, over most
of the MSSM parameter space, at most one Higgs boson would be detectable at the
LHC or LEP. Since Higgs boson couplings cannot be measured with any precision
at hadron colliders, it would be
very difficult to distinguish the SUSY and SM Higgs sectors.
In contrast, at a 500~GeV Linear Collider, the discovery of one of the MSSM
Higgs bosons is guaranteed with just 1 $fb^{-1}$ of integrated luminosity.
Furthermore, if $m_{H_p} \leq 400$~GeV,
with an integrated luminosity of 50 $fb^{-1}$,
it should be possible to distinguish\cite{JANOT}
the MSSM and SM Higgs sectors, either
by directly identifying more than one of the Higgs bosons,
or by detecting a measurable deviation
of the branching fraction for the decay $H_{\ell} \rightarrow b\bar{b}$ from
its SM expectation.

\subsection{Supergravity Phenomenology}
Supergravity (SUGRA) GUT models, via specific assumptions about the symmetries
of interactions responsible for SUSY breaking, provide an economical
framework for phenomenology by relating the many SUSY
breaking parameters of the MSSM. These relations hold
at some ultra-high unification scale
$M_X$ where these symmetries are manifest and the physics is simple.
Complex sparticle mass and mixing patterns (recently incorporated
into ISAJET 7.10),
{\it along with the correct breaking of electroweak symmetry}
emerge\cite{SPECTRA}
when these parameters are renormalized down to the weak scale as required for
phenomenology. The model is completely specified by just four SUSY
parameters which may be taken to be the common values of
SUSY-breaking gaugino ($m_{1/2}$) and scalar
masses ($m_0$) and trilinear scalar couplings ($A_0$),
all specified\footnote{The reason for the hierarchy between the values of
the soft-breaking parameters, which must be  be $O$(1~TeV),
and the scale $M_X$ at which they are specified
is not well understood. It, presumably, has to do with the unknown
dynamics of supersymmetry breaking. At present, this ratio has to be put in
``by hand'', so that it would be premature to claim that supergravity models
provide an understanding of the weak scale.}
at the scale $M_X$, together
with the Higgs sector parameter $\tan\beta$.
In particular, $\mu$ and
the pseudoscalar Higgs boson mass are determined.
Within this framework, the first two generations of squarks are approximately
degenerate (consistent with FCNC constraints in the K-meson sector) while
some third generation squarks may be significantly lighter. The splitting
between squarks and sleptons is as in Eq. (4). The various flavours of
left- (right-) type sleptons are almost exactly degenerate.
In the gaugino-Higgsino
sector, SUSY-breaking electroweak gaugino
masses are typically considerably smaller than $|\mu|$,
so that the $\tw_1,\tz_1$
and $\tz_2$ are dominantly gaugino-like. This has important repercussions
for the cascade decay patterns of heavy sparticles.
It is remarkable that even the simplest SUGRA GUTs are
consistent with experimental constraints as well as
cosmology\cite{SUGRA}.

Since the phenomenology is determined in terms of just
four SUSY parameters, various SUSY
cross sections become
correlated. Various analyses
from LEP and Tevatron experiments
can thus be
consistently combined into a single framework\cite{BCMPT},
as illustrated in Fig. 4 for (a) $\mu < 0$ and (b) $\mu > 0$.
Here, we have taken $A_0 = 0$ (this does not mean that the weak scale
$A$-parameter vanishes) and $\tan\beta =2$, and performed the analysis in the
$m_0-m_{1/2}$ plane.
\begin{figure}[bt]
\vspace{135mm}
\caption {Excluded regions of the $m_0-m_{1/2}$ plane,
together with the reach of the Tevatron and LEP II
within the framework of the minimal supergravity model with radiative
electroweak symmetry breaking. The other parameters are as stated in the text.
The shaded regions are excluded by
theoretical considerations while the region below the hatched lines are not
allowed by current experimental constraints. The dashed and dash-dotted
lines denote the reach of the Tevatron and LEP II, respectively, as discussed
in the text. This figure is taken from Ref.[76] to which we refer
the reader to for further details.}
\end{figure}
The shaded region
is excluded as it either does not lead to the correct EWSB pattern,
or yields a sparticle other than $\tz_1$ as the LSP.
The region below the hatched line is excluded by experimental constraints.
It comprises of the regions excluded by the non-observation of Higgs bosons,
the chargino or sleptons at LEP, or, below the line labelled $\eslt$, by
the gluino and squark search at the Tevatron. For details, we refer the
reader to Ref.\cite{BCMPT} from which this figure is taken. The dot-dashed
lines show the projected reach of LEP II, whereas the dashed line
labelled 200 $fb$ (20 $fb$) indicates
the reach of the Tevatron to the trilepton signal for an integrated
luminosity of 100 $pb^{-1}$ (1 $fb^{-1}$). Finally, the dotted lines
indicate the boundaries of the region where the various two body decays of
$\tz_2$ become kinematically accessible.
It is interesting to see
that since the various
searches frequently probe different parts of the parameter space
they often complement one another. For large values of $A_0$, top squark
signals also play a role in probing the parameter-plane in the Fig. 4.
It should, however, be remembered
that this framework depends on assumptions about physics at the unification
scale. These assumptions, of course, need to be subjected to experimental
tests. These unification hypotheses are directly falsifiable by precision
measurements that will be possible at Linear Colliders.
For current experimental analyses we
suggest using SUGRA models to obtain
default values of MSSM input parameters, and then, to test the sensitivity of
the predictions on the assumed SUGRA relations.

The simplest SUGRA GUT models make several striking predictions. First,
there is the unification of gaugino masses (this has to do with GUTs and not
SUGRA) that we have already discussed. Second, the assumed unification of
scalar masses together with our assumption regarding the minimality of the
sparticle content, implies definite relations amongst the masses of the
unmixed squarks and sleptons. These can be directly verified\footnote{This
is especially important since an alternative proposal
for a high energy symmetry
to suppress FCNC does not require squarks to be degenerate\cite{SEIB}.}
since it should
be possible to precisely measure squark\cite{FF}
and slepton\cite{MUR} masses at the
*LC. Somewhat more detailed tests that make use of the availibility of
polarized beams may also be possible. For example, if $\te_R$ and the lighter
chargino are both kinematically accessible, the measurement of $m_{\te_R}$,
$m_{\tz_1}$,
$m_{\tw_1}$, $\sigma_R$($\te_R$) and $\sigma_R$($\tw_1$), would allow
one to make a global fit and so determine
the two electroweak gaugino masses, $\mu$ and $\tan\beta$ (with a precision
depending on where we are in parameter space). This allows one to test
the gaugino unification condition at the 5\% level\cite{MUR}. The
determination of the other parameters would serve to restrict other SUSY
reaction rates, and thus, provide further tests of this framework. Observation
of sparticles would not only be a spectacular new discovery, but a measurement
of their properties, particularly at linear colliders
would test various
SUGRA assumptions, and so, serve as a telescope to the unification scale.

\section{Optional New Physics}

As we discussed in Sec. 1, there are many other extensions of the SM than
the ones that we have, motivated by our discussion of EWSB, chosen to focus
on. For instance, it is possible to consider models with larger gauge groups,
with the additional associated gauge bosons. Because these groups usually
have large representations, such models frequently include additional matter
fermions as well as additional spin zero bosons. The new physics spectrum
depends on the essentially unknown scale at which the larger symmetry
reduces to the SM gauge group. There already exist experimental limits
on the masses of these exotic particles. The Tevatron experiments have
searched for new, heavy $W$ and $Z$ bosons via their leptonic decays. Assuming
that these have the same couplings as the SM gauge bosons, they obtain direct
bounds\cite{WOOD} $M_{W'} \geq 620$~GeV, $M_{Z'} \geq 500$~GeV on their
masses. Lower limits of $\sim$80-120~GeV have
also been obtained for leptoquarks
and coloured stable (or long-lived) exotics. LEP experiments give lower
limits $\sim \frac{M_Z}{2}$ on new leptons, or excited lepton states that
may exist in composite models. These experiments also constrain\cite{POHL}
new $Z'$ bosons that mix with the SM $Z$. We refer the reader to the
literature for details on these and other issues.

\section{Epilogue}

Over the years, we have become used to hearing about the spectacular
successes of the SM. As we have heard at this meeting, there is no glaring
experimental discrepancy, though some measurements could possibly be taken
as hints of a problem. Most theorists, however, regard the SM as an
incomplete picture since it leaves many things unexplained.
While there have been many ambitious attempts to write down ``Theories of
Everything'' these frequently suffer from the fact that dynamical
calculations are not possible with present techniques, so that it is
difficult to test the underlying ideas.
Quite possibly, progress on the various items
on the theorist's wish-list may occur one step at a time. The problem is
that there is very little guidance from experiment as to the best strategy
for extending the SM.

The one clue we have comes from the
faith that the EWSB sector should not require uncanny fine-tuning, order by
order in perturbation theory. It is possible that the
fine-tuning issue may turn out to be an artifact of our theoretical
techniques, but this is the only guidance that we have. If we take it
seriously,
general arguments suggest that there must be some new degrees of freedom
not included in the SM, that will manifest themselves in elementary particle
collisions at the TeV scale. These arguments, however, tell us nothing about
the nature of the new physics.

We have examined collider signals from two broadly different
extensions of the SM. In the first case, it is assumed that EWSB
interactions become strong at the TeV scale, and that a fermion condensate
is responsible for symmetry breaking. Extended Technicolour Models are an
example. The observation of new bound states, the
Techni-particles, would provide direct confirmation of the this idea.
It could, however, be that these states are kinematically inaccessible at
colliders. In this case, the detection of the signal
will pose a formidable
challenge, both at the LHC as well as at a 1.5-2~TeV $e^+e^-$ collider.
The best hope of accessing the new physics is
via a study of longitudinal $W$ and $Z$
boson scattering at high energy. The modification of the
scattering amplitude due to the new strong interactions of gauge bosons
which serves as the signal
is somewhat sensitive to the details of the model. In many cases, the signal
is rather marginal.\footnote{Since the signal to background ratio
improves dramatically with energy,
a Large Super Giant New Accelerator (LSGNA) would be the optimal facility
for an in depth study of
such a scenario. Presumably, the type of machine and the design
parameters will be
decided after an assessment of
what is found at the next generation of accelerators.}
The
LHC with an integrated luminosity of $\sim 200 \ fb^{-1}$ has roughly
similar capability as a 1.5-2~TeV $e^+e^-$ supercollider with a
similar integrated
luminosity .

The only known realization of the second case, where the EWSB sector is weakly
coupled, is TeV scale supersymmetry.
Unlike as in the case of Technicolour, it is possible to construct
consistent and calculable models in agreement with phenomenology. Supergravity
GUT models are particularly attractive in that they have relatively few
additional parameters. Although this is not absolutely compulsory,
the electroweak gauginos are typically only a third as heavy as gluinos
in the simplest versions of SUSY GUT models.
Thus an $e^+e^-$ collider operating at an energy
$\sim$1~TeV would have a similar reach (via charginos) as the LHC in gluinos.
But there may be a much more significant sense in which
the LHC and the *LC may be complementary.\footnote{For a more detailed
discussion of the complementarity of these facilities, see Ref.\cite{INS}.}
At the LHC, it is likely that
it will be possible to detect gluinos in both the $\eslt$ as well as
multilepton channels, which will serve as indirect evidence for the existence
of charginos and neutralinos. The detailed properties of charginos and
neutralinos will, however, be measured only at Linear Colliders. Re-analysis
of the LHC data in light of this new information, could then help sort out the
complicated cascade decay patterns, which would otherwise be very difficult
to disentangle.
The complementarity of $e^+e^-$ and hadron colliders is also clear if we
consider the search for Higgs bosons. An $e^+e^-$ collider operating at
500~GeV would be an ideal facility for detecting intermediate mass Higgs
bosons which are generally difficult to detect at the LHC. Furthermore, at
Linear Colliders, it
is frequently
possible to distinguish the SM Higgs sector from that of more complicated
models.

We conclude by underscoring the necessity of experimentally
exploring the TeV scale in order to obtain clues about some of the most
pressing questions in particle physics.
While no one knows what we will find, we are
virtually guaranteed to find {\it something}, assuming of course that the
detectors function as advertised. Hopefully, these discoveries will shed
light on some of the items on the theorist's wish list we introduced in
Section 1. Perhaps, we will discover that the grand desert is populated
with unexpected surprises; perhaps, we will find that the new physics
directly probes the dynamics at ultra-high energy scales as, for instance,
in SUGRA GUTs. One thing is certain: we
will find nothing unless we look.

\section{Acknowledgements}

I thank the organizers for their invitation to the conference, and for their
efforts in providing a stimulating atmosphere during the meeting.
It is a pleasure to thank H.~Baer, M.~Bisset,
C-H.~Chen, M.~Drees, J.~Feng, R.~Godbole, J.~Gunion, T.~Han, C.~Kao, T.~Kamon,
R.~Munroe, M.~Nojiri, F.~Paige, S.~Pakvasa,
M.~Paterno, M.~Peskin, S.~Protopopescu,
T.~Rizzo, J.~Sender, A.~White and J.~Woodside and
numerous other colleagues for collaborations and discussions on issues
related to this talk. This
research was supported by the U.S. Dept. of Energy grant DE-FG-03-94ER40833.


\begin{thebibliography}{99}
\bibitem{OLCH} A.~Olchevski, {\it these proceedings}.
\bibitem{FERO} M.~Fero, {\it these proceedings}.
\bibitem{PESKIN} M.~Peskin and T. Takeuchi,
Phys. Rev. {\bf D46}, 381 (1992).
\bibitem{LANG} P.~Langacker, {\it these proceedings}.
\bibitem{BURGESS} C.~P.~Burgess {\it et. al.}, {\em Phys. Lett.} {\bf B326},
276
(1994).
\bibitem{KUHLMANN} S.~Kuhlmann, {\it these proceedings}.
\bibitem{SHOCHET} M.~Shochet, {\it plenary talk at the Eighth DPF meeting,
Albuquerque, NM, Aug. 1994.}
\bibitem{LSND} R.~A.~Reeder, {\it presented at the Eighth DPF meeting,
Albuquerque, NM, Aug. 1994.}
\bibitem{KAM} Y.~Fukuda {\it et. al.}, ICRR-preprint 321-94-16 (1994).
\bibitem{ORTHO} J.~Nico, D.W~Gidley, A.~Rich and P.~W.Zitzewitz,
{\em Phys. Rev. Lett.} {\bf 65}, 1344 (1990).
\bibitem{UNIF} U.~Amaldi, W.~de~Boer and H. F\"urstenau, {\em Phys. Lett.} {\bf
B260}, 447 (1991); J.~Ellis, S.~Kelley and D.~Nanopoulos, {\em Phys. Lett.}
{\bf
B260}, 131 (1991); P.~Langacker and M.~Luo, {\em Phys. Rev.} {\bf D44}, 817
(1991).
\bibitem{BARYON} For text book expositions, see
G.~G.~Ross, {\it Grand Unified Theories},
Benjamin-Cummings Publishing Co. Inc. (1985);
R.~Mohapatra, {\it Unification and Supersymmetry}, Second Edition,
Springer-Verlag (1992).
\bibitem{SUZ} A.~Suzuki and J. Panman, {\it these proceedings}.
\bibitem{BETA} L. Zanotti {\it these proceedings}.
\bibitem{KOLB} E.~Kolb and M.~Turner, {\it The Early Universe},
Addison-Wesley (1990).
\bibitem{PERT} D.~Dicus and V.~Mathur, {\em Phys. Rev.} {\bf D7}, 3111 (1973);
B.~Lee, C.~Quigg and H.~Thacker, {\em Phys. Rev.} {\bf D16}, 1519 (1977).
\bibitem{SELF} N.~Cabibbo, L.~Maiani, G.~Parisi and R.~Petronzio, {\em Nucl.
Phys.} {\bf B158}, 295 (1979);
M. Lindner, {\em Z. Phys.} {\bf C31}, 295 (1986).
\bibitem{SUSYHIG} H. Haber and M.~Sher, {\em Phys. Rev.} {\bf D35}, 2206;
M.~Drees, {\em Int. J. Mod. Phys.} {\bf A4}, 3635 (1989); G.~L.Kane, C.~Kolda
and J.~Wells, {\em Phys. Rev. Lett.} {\bf 70}, 2686 (1993). See also, M.~Sher,
William and Mary preprint, 94-07 (1994).
\bibitem{TECH} S.~Weinberg, {\em Phys. Rev.} {\bf D19}, 1277 (1979);
L.~Susskind,
{\em Phys. Rev.} {\bf D20}, 2619 (1979).
\bibitem{SCHW} J. Schwinger, {\em Phys. Rev.} {\bf 125}, 397 (1962).
\bibitem{ETC} S.~Dimopoulos and L.~Susskind, {\em Nucl. Phys.} {\bf B155},
237 (1979); E.~Eichten and K.~Lane, Phys. Lett. {\bf B90}, 125 (1980).
\bibitem{TECHREV} For a general review, see
R.~Kaul, {\em Rev. Mod. Phys.} {\bf 55}, 449 (1983).
\bibitem{HOLDOM} B. Holdom, {\em Phys. Lett.} {\bf B105}, 301 (1985);
K.~Yamawaki,
M.~Bando and K.~Matumoto, {\em Phys. Rev. Lett.} {\bf 56}, 1335 (1986);
T.~Appelquist, D.~Karabali and L.~Wijewardhana, {\em Phys. Rev. Lett.} {\bf
57},
957 (1986).
\bibitem{APPEL} T. Appelquist and J.~Terning, {\em Phys. Lett.} {\bf B315}, 139
(1993); N.~J. Evans, S.~F~.King and D.~A.~Ross, {\em Phys. Lett.} {\bf B303},
295
(1993); N.~J. Evans and D.~A.~Ross, {\em Nucl. Phys.} {\bf B417}, 151 (1994).
\bibitem{CHIV} R.~S.~Chivukula, S.~Selipsky and E.~Simmons,
{\em Phys. Rev. Lett.} {\bf 69}, 575 (1992).
\bibitem{GATES} R.~S.~Chivukula, E.~Gates, E.~Simmons and J.~Terning,
{\em Phys. Lett.} {\bf B311}, 157 (1993).
\bibitem{BAGGER} J.~Bagger {\it et. al.}, {\em Phys. Rev.} {\bf D49}, 1246
(1994).
\bibitem{HI} K.~Hikasa and K.~Igi, {\em Phys. Lett.} {\bf B261}, 285 (1991)
and {\em Phys. Rev.} {\bf D48}, 3055 (1993).
\bibitem{CG} M.~Chanowitz and M.~Gaillard,
{\em Nucl. Phys.} {\bf B261}, 379 (1985).
\bibitem{HAN} T.~Han, in {\it Proc. of Workshop on Physics and Experiments
with Linear $e^+e^-$ Colliders}, Waikaloa, Hawaii, April 1993, F.~Harris,
S.~Pakvasa, S.~Olsen and X.~Tata, Editors, World Scientific (1993).
\bibitem{BACK} E.~Boos and G.~Jikia, {\em Phys. Lett.} {\bf B275}, 164 (1992);
M.~Herrero and E.~Ruiz-Morales, {\it ibid.} {\bf B296}, 397 (1992);
A.~Abbasabadi, D.~Bowser-Chao, D.~Dicus and W.~Repko, {\em Phys. Rev.}
{\bf D49},
1265 (1994); G.~Jikia, {\em Phys. Lett.} {\bf B298}, 224 (1993);
M.~Berger, {\em Phys.
Rev.} {\bf D48}, 5121 (1993); D.~Dicus and C. Kao, {\it ibid.} {\bf D49},
1265 (1994).
\bibitem{BRODSKY} S.~J.~Brodsky, in {\it Proc. of Workshop on Physics and
Experiments
with Linear $e^+e^-$ Colliders}, Waikaloa, Hawaii, April 1993, F.~Harris,
S.~Pakvasa, S.~Olsen and X.~Tata, Editors, World Scientific (1993).
\bibitem{CHEUNG} K.~Cheung, {\em Phys. Lett.} {\bf B323}, 85 (1994).
\bibitem{JIK} G.~Jikia, {\it presented at the Workshop on Photon-Photon
Colliders, Berkeley, March, 1994}.
\bibitem{SUSYREV} For reviews of weak scale supersymmetry phenomenology, see
H.~P.~Nilles, {\em Phys.~Rep.} {\bf 110}, 1 (1984);
H. Haber and G. Kane, {\em Phys.~Rep.} {\bf 117}, 75 (1985);
X.~Tata, in {\it The Standard Model and Beyond},
p.~304, edited by J.~E.~Kim, World Scientific (1991);
R.~Arnowitt and P.~Nath, {\it Lectures presented at the VII J.~A.~ Swieca
Summer School, Campos do Jordao, Brazil, 1993} CTP-TAMU-52/93;
{\it Properties of SUSY
Particles}, L. Cifarelli and V.~Khoze, Editors, World Scientific (1993).
\bibitem{GIRARDELLO} L.~Girardello and M.~Grisaru, {\em Nucl. Phys.}
{\bf B194}, 65 (1982).
\bibitem{RGE} K. Inoue, A.~Kakuto, H.~Komatsu and H.~Takeshita,
{\em Prog. Theor. Phys.} {\bf 68}, 927 (1982) and {\bf 71}, 413 (1984).
\bibitem{DM} See {\it e.g.} H.~Baer, M.~Drees and X.~Tata,
{\em Phys. Rev.} {\bf D41}, 3414 (1990).
\bibitem{RPARITY} For phenomenological reviews of signals in $R$-parity
violating models, see {\it e.g.} D.~P.~Roy, {\it Proc. of the Tenth DAE
Symposium on High Energy Physics}, S.~Banerjee and P.~Roy, Editors;
H.~Dreiner, in {\it Properties of SUSY
Particles}, Ref.\cite{SUSYREV}.
\bibitem{DT} J.~Ellis, S.~Ridolfi and F.~Zwirner, {\em Phys. Lett.} {\bf B237},
423 (1990); M.~Drees and X.~Tata, {\em Phys. Rev.} {\bf D43}, 2971 (1991).
\bibitem{LEP} See {\it e.g.} G. Giacomelli and P.~Giacomelli,
{\em Riv.Nuovo Cim.}
{\bf 16}, 1 (1993).
\bibitem{CAS} H. Baer, J.~Ellis, G.~Gelmini, D.~Nanopoulos and X.~Tata,
{\em Phys. Lett.} {\bf B161}, 175 (1985); G.~Gamberini, {\em Z. Phys.}
{\bf C30}, 605
(1983); H.~Baer, V.~Barger, D.~Karatas and X.~Tata, {\em Phys. Rev.} {\bf D36},
96
(1987).
\bibitem{ISAJET} F. Paige and S. Protopopescu, in {\it Supercollider Physics},
p. 41, ed.\ D. Soper (World Scientific, 1986);
H. Baer, F. Paige, S. Protopopescu and X. Tata, in
{\it Proceedings of the Workshop on Physics at Current Accelerators
and Supercolliders}, ed.\ J. Hewett, A. White and D. Zeppenfeld,
(Argonne National Laboratory, 1993).
\bibitem{TEVGL}  M.~Paterno, Ph.D. Thesis; D.~Claes,
{\it presented at the Eighth DPF meeting,
Albuquerque, NM, Aug. 1994.}
For the published CDF bound, see F.~Abe {\it et. al.},
{\em Phys. Rev. Lett.} {\bf 69} (1992) 3439.
\bibitem{WELLS} J.~Wells, C.~Kolda and G.~Kane, University of Michigan
preprint, UM-TH-94-23 (1994).
\bibitem{TOP} F.~Abe {\it et. al.}, Fermilab preprint Fermilab-Pub-94/097-E
(1994).
\bibitem{CASHMORE} R. Casmore {\it et. al.}, {\em Phys. Rep.} {\bf 122}, 275
(1985).
\bibitem{DREINER} J.~Butterworth and H.~Dreiner, {\em Nucl. Phys.} {\bf B397},3
(1993).
\bibitem{KAMON} T.~Kamon, J.~Lopez, P. McIntyre and J.~White, Texas A and M
preprint, CTP-TAMU-19/94 (1994).
\bibitem{BKTGL} H.~Baer, C.~Kao and X.~Tata, {\em Phys. Rev.} {\bf D48}, R2978
(1993).
\bibitem{RESW} D.~Dicus, S.~Nandi and X.~Tata, {\em Phys. Lett.} {\bf B129},
451
(1983); H.~Baer and X.~Tata, {\em Phys. Lett.} {\bf 155B}, 278 (1985); H.~Baer,
K.~Hagiwara and X.~Tata, {\em Phys. Rev. Lett.} {\bf 57}, 294 (1986) and
{\em Phys. Rev.} {\bf D35}, 1598 (1987); R.~Arnowitt, A. Chamesiddine and
P.~Nath,
{\em Phys. Rev.} {\bf D35}, 1085 (1987).
\bibitem{NA} P.~Nath and R.~Arnowitt, {\em Mod. Phys. Lett} {\bf A2}, 1113
(1987).
\bibitem{BT} H. Baer and X.~Tata, {\em Phys. Rev.} {\bf D47}, 2739 (1993).
\bibitem{NOSCALE}A.~Lahanas and D.~Nanopoulos, {\em Phys. Rep.} {\bf 145}, 1
(1987).
\bibitem{WINO} CDF Collaboration, submitted to {\em 27 International
Conference on High Energy Physics}, Glasgow, Scotland,
Fermilab-Conf-94/149-E (1994); D0 Collaboration,
presented at {\em 9th Topical Workshop on $\bar{p}p$ Collider
Physics}, Tsukuba, Japan, Fermilab-Conf-94-036-E (1994).
\bibitem{BKTWIN} H. Baer, C. Kao and X. Tata, {\em Phys. Rev.}
{\bf D48} (1993) 5175.
\bibitem{ZICH} J.~Lopez, D.~Nanopoulos, X.~Wang and A.~Zichichi, {\em Phys.
Rev.} {\bf D48}, 2062 (1993).
\bibitem{HK} K.~Hikasa and M. Kobayashi, {\em Phys. Rev.} {\bf D36} (1987)
724.
\bibitem{BST} H.~Baer, J.~Sender and X.~Tata, Hawaii preprint UH-511-788-94
(1994), {\em Phys. Rev.} {\bf D} (in press).
\bibitem{GUN} H.~Baer {\it et. al.}, {\em Phys. Rev.} {\bf D44} (1991) 725.
\bibitem{LHC} C.~Albajar {\it et. al.}, {\em Proc. ECFA LHC Workshop}, Aachen
(1990).
\bibitem{ATLAS} ATLAS Collaboration, {\it Letter of Intent,}
CERN LHCC/92-4 (1992).
\bibitem{BTW} H.~Baer, X.~Tata and J.~Woodside, {\em Phys. Rev.} {\bf D45}
(1992) 145.
\bibitem{BCPT} H.~Baer, C-H. Chen, F.~Paige and X.~Tata, {\em Phys. Rev.}
{\bf D} (in press).
\bibitem{SLEP} H.~Baer, C-H. Chen, F.~Paige and X.~Tata, {\em Phys. Rev.}
{\bf D49} (1994) 3283.
\bibitem{DN} M.~Drees and M.~Nojiri, {\em Phys.Rev.} {\bf D49} (1994) 4595.
\bibitem{MUR} T.~Tsukamoto {\it et. al.}, KEK preprint 93-146 (1993).
\bibitem{FENG} J.~Feng and M.~Strassler, SLAC preprint, SLAC-PUB-6497 (1994).
\bibitem{INS} M.~Peskin, {\it Invited Talk, 22nd INS Symposium on Physics
with High Energy Colliders, Tokyo, March, 1994}, SLAC-PUB-6582 (1994).
\bibitem{JON} J.~Feng,  {\it presented at the Eighth DPF meeting,
Albuquerque, NM, Aug. 1994.}
\bibitem{JANOT} P.~Janot in {\em Proc. of the Workshop on Physics and
Experiments with Linear Colliders}, Waikaloa, Hawaii, F.~Harris, S.~Olsen,
S.~Pakvasa and X.~Tata, Editors (World Scientific, 1993).
\bibitem{HIGREV} V.~Barger, M.~Berger, A.~Stange and R.~Phillips, {\em Phys.
Rev.} {\bf D45}, 4128 (1992);
H.~Baer, M.~Bisset, C.~Kao and X.Tata, {\em Phys. Rev.}
{\bf D46}, 1067 (1992); J.~Gunion and L.~Orr, {\em Phys. Rev.} {\bf D46},
2052 (1992); Z.~Kunszt and F.~Zwirner, {\em Nucl. Phys.}
{\bf B385}, 3 (1992); for a recent review,
see {\it e.g.} J.~Gunion in {\it Properties of SUSY
Particles}, Ref.\cite{SUSYREV}.
\bibitem{HIGB} T.~Garvaglia, W.~Kwong and D-D.~Wu, {\em Phys.Rev.}
{\bf D48}, 1899 (1993); J.~Dai, J.~Gunion and R.~Vega, {\em Phys. Lett.}
{\bf B315}, 355 (1993).
\bibitem{SPECTRA} Some recent analyses of supergravity mass patterns include,
G. Ross and R.~G.~Roberts, {\em Nucl. Phys.} {\bf B377}, 571 (1992);
R.~Arnowitt and
P.~Nath, {\em Phys. Rev. Lett.} {\bf 69}, 725 (1992);
M.~Drees and M.~M.~Nojiri,
{\em Nucl. Phys.} {\bf B369}, 54 (1993); S.~Kelley {\it et. al.},
J.~Lopez, D.~Nanopoulos, H.~Pois and K.~Yuan,
{\em Nucl. Phys.} {\bf B398}, 3 (1993);
M. Olechowski and S. Pokorski, {\em Nucl. Phys.} {\bf B404}, 590 (1993);
V.~Barger, M.~Berger and P.~Ohmann, {\em Phys. Rev.} {\bf D49}, 4908 (1994);
G. Kane, C.~Kolda, L.~Roszkowski and J.~Wells, {\em Phys. Rev.} {\bf D49}, 6173
(1994); D.~J.~Casta\~no, E.~Piard and P.~Ramond, {\em Phys. Rev.}
{\bf D49}, 4882 (1994);
W.~de~Boer, R.~Ehret and D.~Kazakov, Karlsruhe preprint,
IEKP-KA/94-05 (1994).
\bibitem{SUGRA} R.~Arnowitt and P.~Nath, {\em Phys. Rev. Lett.} {\bf 69},
725 (1992); J.~Hisano, H.~Murayama and T.~Yanagida, {\em Nucl. Phys.} {\bf
B402}, 46 (1993); J.~Lopez, D.~Nanopoulos and H.~Pois, {\em Phys. Rev.} {\bf
D47}, 2468 (1993);
H.~Baer, M.~Drees, C.~Kao, M.~Nojiri and X.~Tata,
{\em Phys. Rev.} {\bf D50}, 2148 (1994);  J. Lopez,
D.~Nanopoulos, G.~Park, X.~Wang and A.~Zichichi, {\em Phys. Rev.} {\bf D50},
2164 (1994).
\bibitem{BCMPT} H.~Baer, C-H.~Chen, R.~Munroe, F.~Paige and X.~Tata, Hawaii
preprint, UH-511-795-94 (1994).
\bibitem{SEIB} Y.~Nir and N.~Seiberg, {\em Phys. Lett.} {\bf B309}, 337 (1993).
\bibitem{FF} J.~Feng and D.~Finnell {\em Phys. Rev.} {\bf D49}, 2369 (1994).
\bibitem{WOOD} D.~Wood, {\it these proceedings}.
\bibitem{POHL} M. Pohl, {\it plenary talk at the International Conference on
High Energy Physics, Glasgow, August 1994}.
%
\end{thebibliography}
\end{document}